\documentclass[prl,twocolumn,amsmath,amssymb,floatfix,superscriptaddress,nofootinbib,preprintnumbers]{revtex4-1}

\pdfoutput=1
\usepackage{graphicx}
\usepackage{amssymb}
\usepackage{amsmath}
\usepackage{bm}
\usepackage{color}
\usepackage[utf8]{inputenc}
\usepackage{comment}
\usepackage{hhline}
\usepackage{float}
\usepackage[normalem]{ulem}
\usepackage{natbib}

\newcommand{\refeq}[1]{eq.~(\ref{eq:#1})}
\newcommand{\refeqs}[2]{eqs.~(\ref{eq:#1})--(\ref{eq:#2})}

\newcommand{\reffig}[1]{figure~\ref{fig:#1}}
\newcommand{\refFig}[1]{Figure~\ref{fig:#1}}

\newcommand{\cO}{\mathcal O}
\newcommand{\fs}{{\rm fs}}
\newcommand{\dep}{{\rm dep}}

\newcommand{\mnras}{{\em Mon. Not. Roy. Astron. Soc. }}

\definecolor{ultramarine}{rgb}{0.07, 0.04, 0.56}
\definecolor{cadmiumgreen}{rgb}{0.0, 0.42, 0.24}
\definecolor{indigo(dye)}{rgb}{0.0, 0.25, 0.42}
\definecolor{purple}{rgb}{0.75, 0.0, 1.0}
\usepackage[linktocpage=true]{hyperref}
\hypersetup{
colorlinks=true,
citecolor=ultramarine,
linkcolor=cadmiumgreen,
urlcolor=indigo(dye),
pdfauthor={},
pdftitle={},
pdfsubject={}
}

\begin{document}

\preprint{YITP-SB-18-38}

\title{First detection of scale-dependent linear halo bias in $N$-body simulations with massive neutrinos}

\author{Chi-Ting Chiang}

\affiliation{C.N. Yang Institute for Theoretical Physics, Department of Physics \& Astronomy,
Stony Brook University, Stony Brook, NY 11794}

\affiliation{Physics Department, Brookhaven National Laboratory, Upton, NY 11973, USA}

\author{Marilena LoVerde}

\affiliation{C.N. Yang Institute for Theoretical Physics, Department of Physics \& Astronomy,
Stony Brook University, Stony Brook, NY 11794}

\author{Francisco Villaescusa-Navarro}

\affiliation{Center for Computational Astrophysics, Flatiron Institute, 162 5th Avenue, 10010, New York, NY, USA}

\begin{abstract}
Using $N$-body simulations with massive neutrino density perturbations,
we detect the scale-dependent linear halo bias with high significance.
This is the first time that this effect is detected in simulations containing
neutrino density perturbations on all scales, confirming the same finding
from separate universe simulations. The scale dependence is the result of
the additional scale in the system, i.e. the massive neutrino free-streaming
length, and it persists even if the bias is defined with respect to the cold
dark matter plus baryon (instead of total matter) power spectrum. The separate
universe approach provides a good model for the scale-dependent linear bias,
and the effect is approximately $0.25f_\nu$ and $0.43f_\nu$ for halos with
bias of 1.7 and 3.5, respectively. While the size of the effect is small,
it is \emph{not} insignificant in terms of $f_\nu$ and should therefore be
included to accurately constrain neutrino mass from clustering statistics
of biased tracers. More importantly, this feature is a distinct signature
of free-streaming particles and cannot be mimicked by other components of
the standard cosmological model.
\end{abstract}

\maketitle

Massive neutrinos play an important role in cosmology. They contribute
to the energy budget, which impacts the expansion history, and their large
momenta prevent them from  clustering along with cold dark matter (CDM) on
scales smaller than their free-streaming length, $\lambda_{\fs}$. As a
result, the growth of matter fluctuations becomes \emph{scale-dependent},
with fluctuations on sub-free-streaming scales growing more slowly. This
effect suppresses the matter power spectrum for $k > k_{\fs} \sim 1/\lambda_{\fs}$
by an amount proportional to the fraction of the matter budget composed of
neutrinos $f_\nu$ \cite{Eisenstein:1997jh,Hu:1997vi}. Since the neutrino
abundance is fixed by the cosmology and deviations are well-constrained
\cite{Aghanim:2018eyx}, a detection of this effect determines the sum of
the neutrino masses, $\sum m_\nu$. 

Scale-dependent growth of structure from neutrinos also fundamentally
changes the way that the halo (or galaxy or cluster) density field
traces the underlying matter distribution. The relationship between
the two can be parameterized by a bias coefficient $b$, with $\delta_h \sim b \delta_c$,
where $\delta_h$ is the halo number density contrast and $\delta_c$
the CDM+baryon density contrast. In the standard $\Lambda$CDM cosmology,
halo formation is completely local and this dictates that the bias
approaches a constant on large-scales \cite{Kaiser:1984sw, Bardeen:1985tr, 1993MNRAS.262.1065C, Mann:1997df}.
Neutrinos, however, can travel cosmological distances during structure
formation. Consequently, the local gravitational dynamics of particles
forming halos depends upon the history of neutrino density field out
to extremely large scales. This fact allows the halo bias to become
scale-dependent, where the scale introduced is the neutrino free-streaming
scale. Since the bias is itself an observable, this effect offers
another probe of $\sum m_\nu$.

Scale-dependent bias generated by massive neutrinos was first predicted
by Ref.~\cite{LoVerde:2014pxa}, and later measured in $N$-body simulations
using the separate universe technique in Ref.~\cite{Chiang:2017vuk}.
A related effect on the bias of cosmic voids was identified in Ref.~\cite{Banerjee:2016zaa}.
The effect on halos has not yet been detected by any other neutrino simulation
techniques for two main reasons. First, to robustly detect quantities
of $\cO(f_\nu)$, a statistical error bar of $\cO(f_\nu/5)$ is needed.
This is challenging because current constraints on $\sum m_\nu$ correspond
to $f_\nu \lesssim 0.01$. Typically $f_\nu$ is artificially boosted to increase
the amplitude of neutrino effects by increasing the neutrino mass
\cite{Brandbyge:2009ce,Agarwal:2010mt,Bird:2011rb,Hannestad:2011td,Wagner:2012sw,AliHaimoud:2012vj,Inman:2015pfa,Liu:2017now,Bird:2018all,
VillaescusaNavarro:2012ag,Villaescusa-Navarro:2013pva,Castorina:2013wga,Costanzi:2013bha,Castorina:2015bma,Heitmann:2015xma,Yu:2016yfe,Emberson:2016ecv,Adamek:2017uiq,Villaescusa-Navarro:2017mfx},
rather than the number density.
This has the effect of pushing the free-streaming
scale towards the nonlinear scale where there are other, more mundane,
sources of scale dependence that are hard to disentangle. Second,
since the bias effect arises from the sensitivity to neutrino perturbations
at earlier epochs when the comoving free-streaming length was larger,
very large volume simulations are needed. Specifically, for a relatively light
neutrino mass (e.g. $0.05$eV, which gives $\lambda_{\fs}\sim 200$Mpc today)
Gpc-scale simulations are needed. These simulations exist,
but none of them have had the statistical power to detect this effect
\cite{VillaescusaNavarro:2012ag,Villaescusa-Navarro:2013pva,Castorina:2013wga,Costanzi:2013bha,Castorina:2015bma,Heitmann:2015xma,Yu:2016yfe,Emberson:2016ecv,Adamek:2017uiq,Villaescusa-Navarro:2017mfx}.

In this letter, we perform simulations with neutrino density perturbations
that meet the above requirements. The degenerate neutrino mass is set to
$m_\nu=0.05$eV, so that $\lambda_{\fs}$ is large enough and the nonlinear
contamination is limited. This choice also gives a free-streaming scale
consistent with that from the minimal mass of the normal and inverted
mass orderings, and expectations given current cosmological constraints
($\sum m_\nu \le 0.12$eV at 95\%  \cite{Aghanim:2018eyx}). To enhance
$f_\nu$ so that effects can be probed by a reasonable number of simulations,
we increase the number of neutrino species to $N_\nu=28$ to give $f_\nu \approx 0.1$.
\footnote{We choose $f_\nu\approx0.1$ so that the effect is large
enough to be detected by a reasonable amount of simulations, and is small
enough so that the neutrino effects are linear in $f_\nu$.}
We set the size of the simulation box to $L=4000h^{-1}~{\rm Mpc}$ to probe the bias
out to the largest scales. Cosmological parameters are as follows: Hubble
constant $h=0.7$, baryon density $\Omega_b=0.05$, CDM density $\Omega_c=0.25$,
CMB temperature $T_\gamma=2.725$K, helium fraction $Y_{\rm He}=0.24$,
the initial curvature power spectrum with the spectral index $n_s=0.95$,
and the amplitude which sets $\sigma_8=0.83$ today for the power spectrum
of CDM+baryon field.

These simulation choices pose two challenges. First, for such light neutrino
masses, a substantial fraction of neutrinos are relativistic at the simulation
starting redshift (e.g. at $z_i=99$, $\rho_\nu(a_i)/(\rho_{\nu,0}a_i^{-3})\sim1.5$).
Thus, simulation methods that treat neutrinos as non-relativistic particles can
lead to unphysical effects.\footnote{One would have to wait until $z\sim 10$
for half of the neutrino population to be non-relativistic ($p \lesssim 0.1m_\nu$,
say); simulations with late starting redshifts will be inaccurate due to transients
from initial conditions \cite{Crocce:2006ve}, which may well be systematically
different between simulations with and without massive neutrinos.} Second, on
the largest scales, radiation perturbations become important, and ignoring them
will lead to incorrect growth (see, e.g. Ref.~\cite{Zennaro:2016nqo}). To
address these problems, we include neutrino and photon perturbations on grids
in our gravity solver \cite{Brandbyge:2008js}. Concretely, the dynamics of the
CDM+baryon field is carried out by \texttt{Gadget-2} \cite{Springel:2005mi},
and we modify the particle-mesh (PM) potential as
\begin{align}
\label{eq:potential}
 \Phi_{\rm tot}(k,a)\:&=\Phi_{bc}(k,a)\Bigg\lbrace1+2\frac{f_\gamma(a)}{f_{bc}(a)}\frac{T_\gamma(k,a)}{T_{bc}(k,a)} \\ \nonumber
 \:&\hspace{0.5cm}+\frac{f_\nu(a)}{f_{bc}(a)}\frac{T_\nu(k,a)}{T_{bc}(k,a)}\left[1+3c^2_{s,\nu}(a)\right]\Bigg\rbrace \,.
\end{align}
Here, $\Phi_{bc}$ is the CDM+baryon potential computed from the simulation particles,
$f_x=\bar{\rho}_x/(\bar{\rho}_\gamma+\bar{\rho}_\nu+\bar{\rho}_{bc})$, $T_x$
is the linear transfer function computed by \texttt{CLASS} \cite{Blas:2011rf,Lesgourgues:2011rh},
and we approximate the neutrino sound speed $c^2_{s,\nu}(a)\approx\dot{\bar{\rho}}_\nu(a)/\dot{\bar{P}}_\nu(a)$
\cite{Lesgourgues:2011rh,Munoz:2018ajr}. Since baryons are treated as CDM in
simulations, hereafter CDM can refer to CDM+baryons. CDM particles are then
displaced according to $\Phi_{\rm tot}$, instead of $\Phi_{bc}$, to account
for the photon and neutrino perturbations. We repeat the procedure at each
time step for the long-range force calculation. The simulation initial conditions
are set up using the Zel'dovich approximation \cite{Zeldovich:1969sb} with
the CDM+baryon linear power spectrum and the scale-dependent growth rate
computed by \texttt{CLASS} at $z_i=99$. We set the number of particles
sampling CDM+baryon perturbations and PM grids to be $1536^3$ and $4096^3$,
respectively.

The accuracy of this approach relies on two assumptions. First, photon and
neutrino perturbations are neglected on scales smaller than the PM grid size:
$\sim1h^{-1}~{\rm Mpc}$. This is a good approximation because the size of the PM grid
is much smaller than the neutrino free-streaming scale and the photon free-streaming
scale (the Hubble radius) at all redshifts of interest. Second, while linear
photon and neutrino perturbations of mode $k$ can affect the nonlinear evolution
of CDM particles of the same $k$, nonlinear evolution of neutrino and photon
perturbations is ignored. Relatedly, our calculation of the total potential
assumes that the photon and neutrino perturbations have the same phases as
CDM+baryon perturbations on all scales. By construction, for our neutrino
mass $\lambda_{\fs}\sim 200$Mpc today, which is large in comparison to
the nonlinear scale ($\sim 10$Mpc) and also the Lagrangian radii of halos
($\lesssim10$Mpc). At higher redshift, and for photons, the free-streaming
scales are even larger, so corrections from the nonlinear photon and neutrino
perturbations should be negligible. Moreover, it has been shown in
Refs.~\cite{VillaescusaNavarro:2011ys,Ichiki:2011ue,VillaescusaNavarro:2012ag,LoVerde:2014rxa,Senatore:2017hyk}
that neutrino clustering around halos is negligible as long as the individual
neutrino masses have $m_\nu\lesssim 0.2$eV. To check the performance of our
modification, we compare the CDM+baryon power spectrum computed from simulations
and \texttt{CLASS} at $z=1$ and 0. On linear scales of $1.57\times10^{-3}\le k/(h~{\rm Mpc}^{-1})\le10^{-2}$,
the fractional difference is less than 0.5\%,\footnote{One cannot achieve
this level of agreement between simulations and Boltzmann code if photon
and neutrino perturbations are not included. See e.g. Ref.~\cite{Zennaro:2016nqo}
for the large-scale difference between the full Boltzmann calculation and
the two-fluid approximation without including radiation perturbations.}
which is smaller than the effect that we are targeting.

We identify halos using the Amiga Halo Finder \cite{Gill:2004km,Knollmann:2009pb}.
Since there are no neutrino particles in the simulations, halos are
identified using CDM particles alone. We set the density threshold
to $\Delta=200$, and the minimum number of particles in halos to be
20. We split the dark matter halos between two catalogs:
$2.75\times10^{13}\le M_h/(h^{-1}~M_\odot)<2.75\times10^{14}$ and
$2.75\times10^{14}\le M_h/(h^{-1}~M_\odot)<2.75\times10^{15}$.
The halo bias is defined as
\begin{equation}
 b(k)=P_{hc}(k)/P_{cc}(k) \,,
\end{equation}
where $P_{hc}$ and $P_{cc}$ are respectively the halo-CDM and CDM-CDM power
spectra measured from simulations. This definition greatly reduces the scale
dependence of the bias in comparison with a bias defined with respect to the
total (neutrino and CDM+baryon) matter density field
\cite{Villaescusa-Navarro:2013pva,Castorina:2013wga,LoVerde:2014pxa, Vagnozzi:2018pwo}.
The alternative bias definition is appropriate when combining galaxy and lensing
data and can provide additional constraints on neutrino mass
\cite{LoVerde:2016ahu,Raccanelli:2017kht,Schmittfull:2017ffw}.
In addition to the massive neutrino simulations, we also run simulations with
a reference cosmology \emph{without} (massive or massless) neutrinos. The rest of the procedures
and cosmological parameters are identical to simulations with massive neutrinos.
In particular, photon perturbations are included in the potential in \refeq{potential}
and the scale-dependent CDM+baryon linear power spectrum and growth rate are
used to set the initial conditions.

Note that while we fix the value of $\sigma_8$ of the CDM+baryon power
spectrum to be the same today for the two cosmologies, this does not
make the two power spectra the same since they have different shapes.
For a fixed $\sigma_8$, the massive neutrino cosmology has a larger power
spectrum for $k \ll k_{\fs}$ and a slightly smaller power spectrum on
smaller scales. This difference makes the function $\sigma(M_h)$, the
variance of perturbations smoothed on scale $R\sim (M_h/\rho_c)^{1/3}$,
different at all redshifts, with the massive neutrino cosmology having
a larger amplitude $\sigma(M_h)$ at the high-mass end and smaller $\sigma(M_h)$
at the low mass end.\footnote{The absence of neutrinos contributing
to the global energy density also changes the expansion history, which
changes the redshift evolution of $\sigma_8(z)$, $\sigma(M_h,z)$, but
this effect is subdominant.} As a result, halos of the same mass will
have different abundances and different bias factors in each cosmology.

\begin{figure}[hp]
\centering
\includegraphics[width=0.48\textwidth]{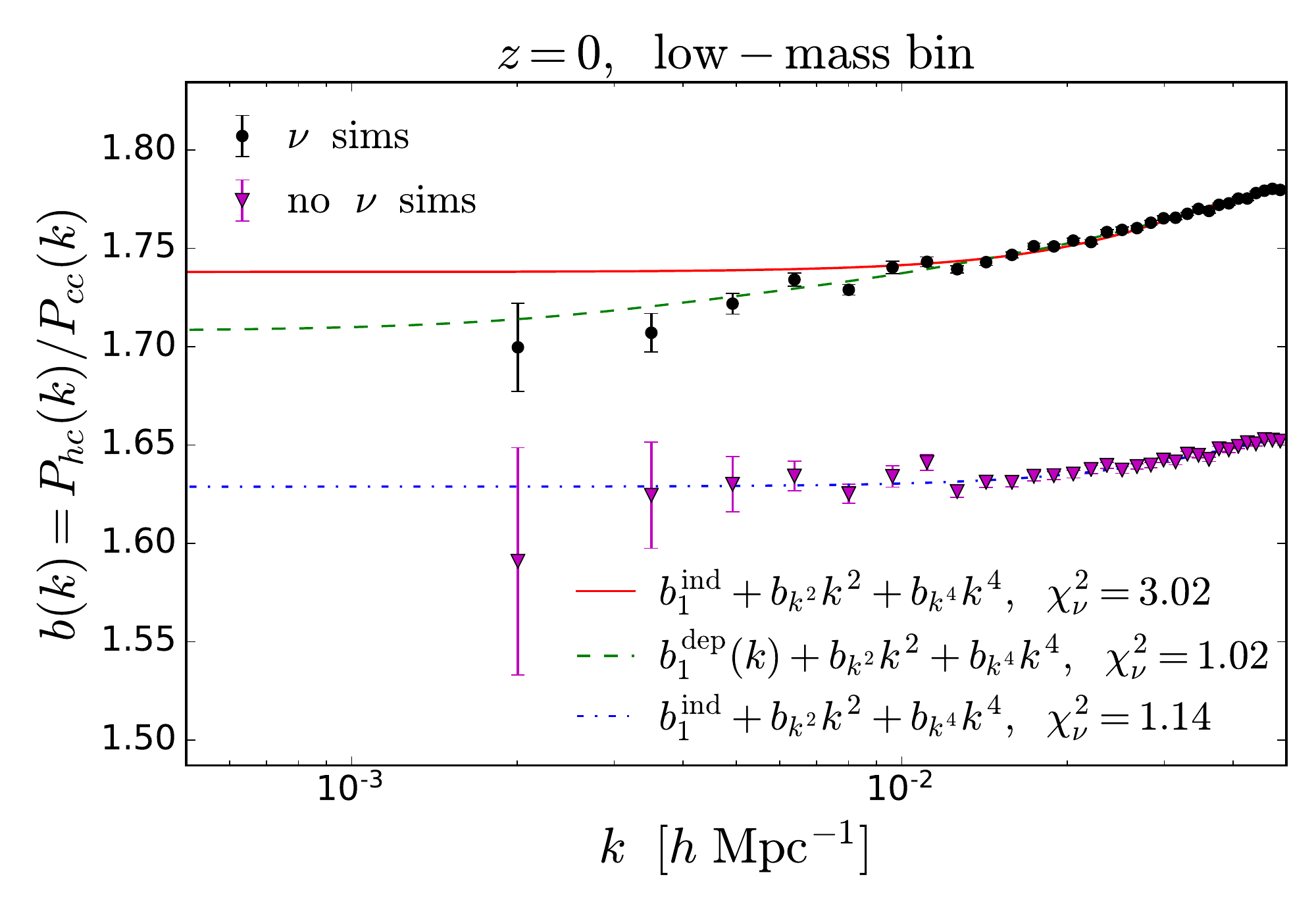} 
\includegraphics[width=0.48\textwidth]{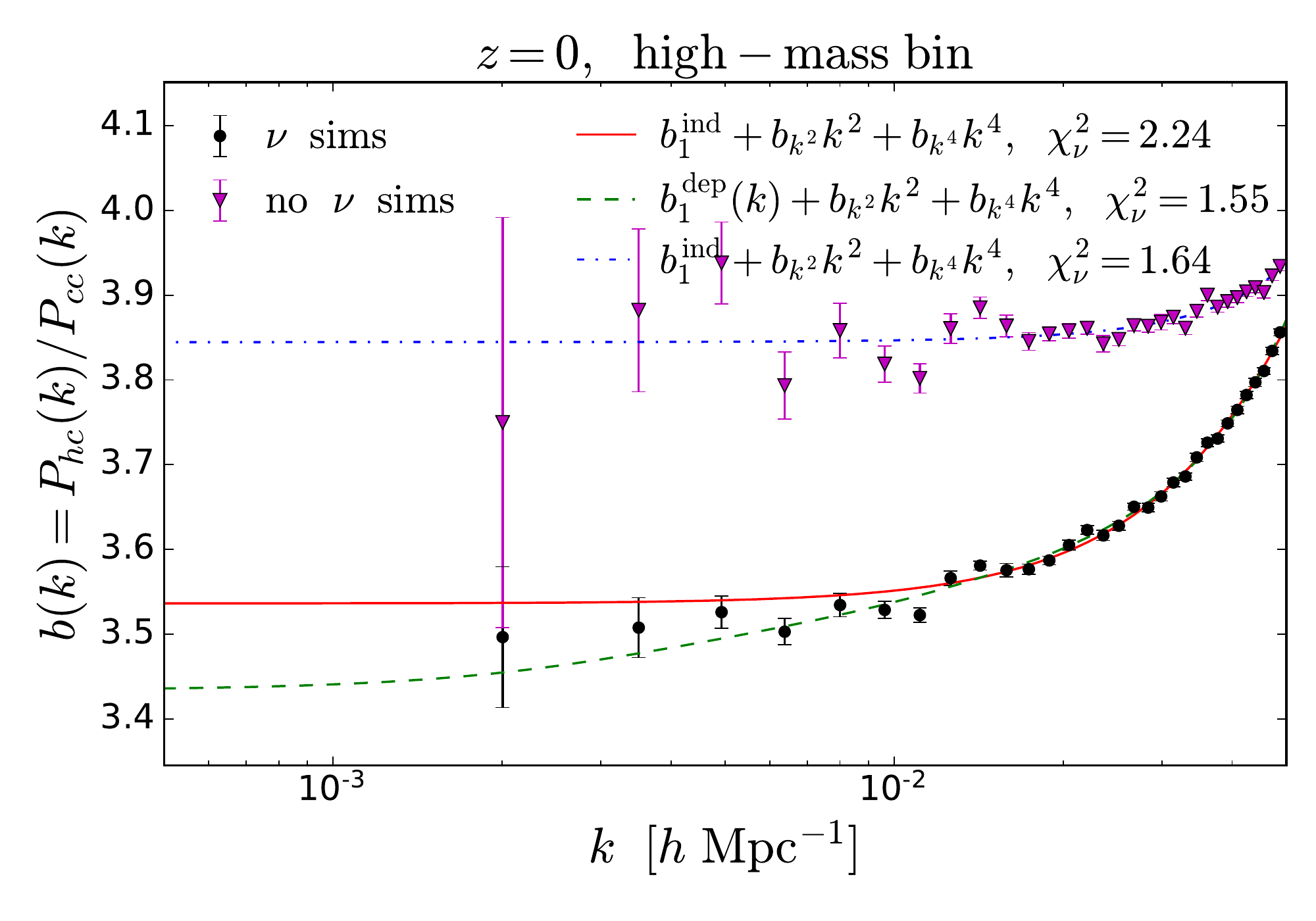} 
\includegraphics[width=0.48\textwidth]{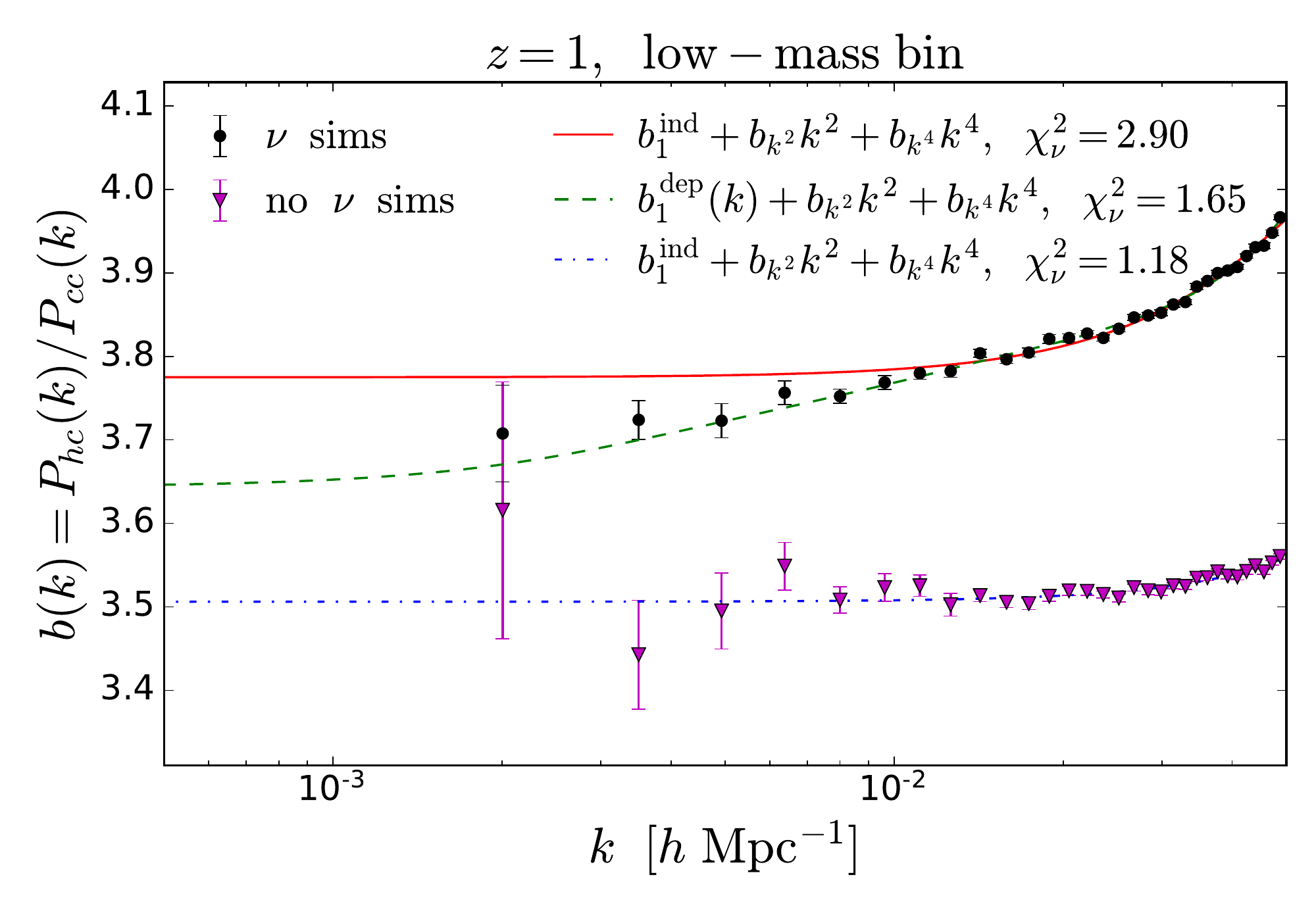} 
\caption{Halo bias measured from 40 simulations with (black circles)
and without massive neutrinos (magenta triangles). The errors bars
show the error on the mean. From top to bottom: low-mass bin at $z=0$,
high-mass bin at $z=0$, low-mass bin at $z=1$. Each panel shows the
best fits for the bias in the massive neutrino simulations using the
scale-independent (red solid) and scale-dependent (green dashed) models
in \refeq{bconst} and \refeq{bk}. The blue dot-dashed lines show the
best-fit bias for the no-neutrino simulations using the scale-independent
model in \refeq{bconst}. The fits are performed within $1.57\times10^{-3}\le k/(h~{\rm Mpc}^{-1})\le 5\times10^{-2}$.
The reduced $\chi^2_\nu$ values shown in the legend are estimated
from \emph{variance only}.}
\label{fig:bk}
\end{figure}

\refFig{bk} shows the halo bias measured from 40 simulations with neutrinos
(black circles) and without neutrinos (magenta triangles). The error bars
show the error on the mean. From top to bottom the panels show the results
for the halo catalogs of a low-mass bin at $z=0$, a high-mass bin at $z=0$,
and a low-mass bin at $z=1$. The error bars are much larger for high-mass
halos at $z=1$ due to their rarity, so we do not show the result. The most
obvious feature in \reffig{bk}, is the difference in the overall amplitude
of the bias between the two cosmologies. As discussed, this is to be expected
from the difference in $\sigma(M_h)$ and we find good agreement with numerical
predictions for halo bias that take $\sigma(M_h)$ as input (e.g. Ref.~\cite{Sheth:1999mn}).
In particular, the relative amplitude of $\sigma(M_h)$ for the two cosmologies
flips between low- and high-mass halos, resulting a change in bias amplitude
seen in \reffig{bk}.
The differences in the errors on the bias are also in agreement
with the analytic calculations that take the stochasticity between the halo
and matter fields to be inversely proportional to the number density of halos,
giving $\sigma_b^2 \propto 1/(nP_{cc})$. For the halos in \reffig{bk},
the difference in the power spectra dominates the difference in the error
bars and our parameter choices give a larger $P_{cc}$ in the massive neutrino
cosmology, resulting in smaller errors on the bias on these scales
for both mass bins.

At present, the amplitude of the bias is not predicted with sufficient accuracy
to extract reliable constraints on neutrino mass so it is usually treated as
a free parameter in cosmological analyses. Given this, we move on to examine
the scale dependence of the bias. At the highest $k$, the bias in both cosmologies
has an upturn consistent with $k^2$ and $k^4$ bias terms 
\cite{Assassi:2014fva,Biagetti:2014pha}, the amplitudes of which are also typically treated as
free parameters that do not constrain cosmology \cite{Giusarma:2018jei}. The most interesting feature
in \reffig{bk}, however, is the persistent difference in the $k$-dependence of
the biases at larger scales (e.g. $k\sim 10^{-3}-10^{-2}h~{\rm Mpc}^{-1}$). These scales
are well into the linear regime and are therefore robust to nonlinear contamination
and baryonic effects. We fit the measurements to two different bias models:
(i) a model with scale-independent linear bias,
\begin{equation}
 b(k)= \underbrace{b_1}_{\equiv b_1^{\rm ind}} + b_{k^2}k^2 + b_{k^4}k^4 \,,
\label{eq:bconst}
\end{equation}
and (ii) a scale-dependent model 
\begin{equation}
 b(k)=\underbrace{1 + (b_1-1) f(k)}_{\equiv b_1^{\dep}(k)}+ b_{k^2} k^2 + k_{k^4}k^4 \,.
\label{eq:bk}
\end{equation}
In both models there are three free parameters, $b_1$, $b_{k^2}$, and $b_{k^4}$,
each are fit separately for every mass bin, redshift, and cosmology. The factor
$f(k)$ in \refeq{bk} is a redshift- and cosmology-dependent function, taken from
the separate universe prediction for how the local power spectrum changes in
response to a long-wavelength CDM+baryon perturbation $\delta_c(k)$ \cite{Chiang:2017vuk}.\footnote{Alternatively,
$f(k)$ can be modeled from the spherical collapse in a region with a long-wavelength
$\delta_c(k)$ \cite{LoVerde:2014pxa,Munoz:2018ajr}, but for massive neutrinos the
two approaches give equivalent predictions \cite{Chiang:2017vuk}.} Importantly,
$f(k)$ does not depend on the mass of the halos. Photon perturbations on horizon
scales also introduce scale dependence to the bias; the physics is identical to
that of massive neutrinos (the presence of a free-streaming length) but the amplitude
is much smaller because of the smaller photon energy density so it can be neglected.
In fitting the parameters in \refeqs{bconst}{bk} we use $1.57\times10^{-3}\le k/(h~{\rm Mpc}^{-1})\le 5\times10^{-2}$,
and we have checked that our fits are insensitive to the precise vale of $k_{\rm max}$.

The red solid and green dashed lines show the best-fit models of scale-independent
and scale-dependent linear bias to the simulations with massive neutrinos,
respectively. The blue dot-dashed line displays the best-fit model of the
scale-independent linear bias for the simulations without neutrinos. In
the legend we show the reduced $\chi^2_\nu$ values from the fit, which
are computed using the \emph{variance only}, as we do not have enough simulations
to precisely measure the covariance matrix. We expect the covariance matrix
to be highly diagonal on scales larger than $0.05h~{\rm Mpc}^{-1}$, but the reduced
$\chi^2_\nu$ values may be slightly underestimated. Thus, these
numbers can serve as a guide to the goodness-of-fit, but they should not be
used to compute $p$-values. We first examine the result for the massive neutrino
simulations. It is clear, both visually and quantitatively through $\chi^2_\nu$,
that for all halo catalogs, the scale-dependent linear bias model is a better
fit to the data than the scale-independent linear model for simulations with
massive neutrinos. We emphasize that even if the halo bias is defined with
respect to the CDM+baryon power spectrum, instead of total matter power spectrum
as suggested by Refs.~\cite{Villaescusa-Navarro:2013pva,Castorina:2013wga},
the scale dependence persists and extends to scales where nonlinearities are
truly negligible. \emph{This is the first time that scale-dependent linear halo
bias is detected in simulations with massive neutrino perturbations}, hence
further confirming the finding from separate universe simulations \cite{Chiang:2017vuk}.
While we do not have measurements on scales larger than $\sim10^{-3}h~{\rm Mpc}^{-1}$,
we nevertheless show the prediction for the scale-dependent bias there to
illustrate that it ultimately approaches a constant with $k$, on scales
where neutrinos have always clustered in the same way as CDM and baryons. 

We next examine the results for the simulations without neutrinos. In all
cases, the halo bias is statistically consistent with being scale independent.
More importantly, as we move to larger scales, the data points become flatter,
suggesting that there are no spurious scale-dependent features on the largest
scale in our simulations.

The presence of the scale-dependent linear halo bias calls for more accurate
modeling to extract neutrino information from the clustering statistics
of biased tracers. The separate universe calculation, which provides a
good model for the halo bias measured from our simulations, predicts that
$b_1^{\dep}(k_\downarrow)/b_1^{\dep}(k_\uparrow)\approx1+0.25f_\nu$ and
$1+0.43f_\nu$ for halo populations with $b_1^{\dep}(k_\uparrow)=1.7$ and
3.5, respectively. Here, $k_\downarrow$ and $k_\uparrow$ denote wavenumbers
in the asymptotic regimes on either side of the neutrino free-streaming scale.
Since the linear halo power spectrum is proportional to $(b_1^{\dep})^2$, the
neutrino effect will be doubled at the leading order compared to the raw bias.
The effect is therefore small, but \emph{not} negligible in terms of $f_\nu$
and there are important consequences. First, since the scale dependence of
the linear bias is opposite to that of the CDM+baryon power spectrum (an
increase, as opposed to suppression at higher $k$), ignoring it can lead
to an underestimation of $f_\nu$. Second, since the size of the effect
depends on both $f_\nu$ and $b_1$, it introduces a degeneracy between
those parameters, which can degrade the constraints on $f_\nu$. On the
other hand, this scale-dependent linear bias is a distinct signature of
free-streaming particles that can exist only if there is an additional
scale in the system. In the standard cosmological model, only massive
neutrinos provide this large-scale characteristic length and so this
signature offers a smoking gun for their detection.

Neutrinos also imprint a characteristic signature on the power spectrum:
the change in the amplitude across the free-streaming scale. Yet detecting
this signature is challenging because cosmic variance fundamentally
limits the precision of measurements of the power spectrum on large scales.
In contrast, measurements of a deterministic linear bias as seen in \reffig{bk}
are not limited by cosmic variance, but by the stochasticity between halos
and the CDM+baryon field, which is thought to be small for halos of very
high number density \cite{Seljak:2008xr}. A detection of scale-dependent
bias is therefore possible with even a small number of Fourier modes
spanning the transition scales in \reffig{bk} \cite{LoVerde:2016ahu}.
Moreover, the dependence of this effect on $b_1$ allows for constraints
on $\sum m_\nu$ to be extracted from the ratio of the bias factors of two
different galaxy populations within a single survey. Surveys that aim to
constrain the bias at horizon-scales to study primordial physics may be
of relevance \cite{Dore:2014cca}.

Finally, we note that the separate universe formalism used to predict
$f(k)$ in \refeq{bk} and \reffig{bk} also predicts a scale-dependent
feature in the squeezed-limit bispectrum \cite{Chiang:2017vuk,Chiang:2016vxa}.
Unfortunately that signal is much smaller and we estimate that roughly
$10\times$ more simulations are needed to confirm a detection.

\section*{Acknowledgments}
\begin{acknowledgments}
Results in this paper were obtained using the Gordon cluster in the San Diego
Supercomputer Center, the Rusty cluster at the Center for Computational Astrophysics,
and the high-performance computing system at the Institute for Advanced Computational
Science at Stony Brook University.
CC was supported by grant NSF PHY-1620628.
ML is supported by DOE DE-SC0017848.
The work of FVN is supported by the Simons Foundation.
\end{acknowledgments}

\bibliographystyle{apsrev4-1}
\bibliography{main}
\end{document}